\documentclass[12pt]{iopart}
\usepackage{graphicx}
 \begin{document}
\title[Nuclear transparency effect of $\pi^{-}$-mesons  M. Ajaz et al.]{Nuclear transparency effect of $\pi^{-}$-mesons in p+$^{12}$C- and d+$^{12}$C- interactions at 4.2A GeV/c}

\author{M. Ajaz$^{1, 2, \dag}$, M. K. Suleymanov$^{1, 3}$, K. H. Khan$^{1}$, A. Zaman$^{1}$}
 \address{$^{1}$COMSATS Institute of Information Technology Islamabad, Pakistan\\ $^{2}$Department of Physics, Abdul Wali Khan University Mardan, Pakistan\\ $^{3}$Veksler and Baldin Laboratory of High Energy Physics, Joint Institute for Nuclear Research Dubna, Russia}
\ead{$^\dag$muhammad.ajaz@cern.ch}

\submitto{\jpg}

\begin{abstract} The use of nuclear transparency effect of $\pi^{-}$-mesons in proton, and deuteron induced interactions with carbon nuclei at 4.2A $GeV/c$, to get information about the properties of nuclear matter, is presented in this work. “Half angle” $(\theta_{1/2})$ technique is used to extract information on nuclear transparency effect. The $\theta_{1/2}$ divides the multiplicity of charged particles into two equal parts depending on their polar angle in the lab. frame in $pp$ interactions. Particles with angle smaller than (incone particles) and greater than (outcone particles) $\theta_{1/2}$ are considered separate. The average values of multiplicity, momentum and transverse momentum of the $\pi^{-}$-mesons are analyzed as a function of a number of identified protons in an event. We observed evidences in the data which could be considered as transparency effect. For quantitative analysis, the results are compared with cascade model. The observed effects are categorized into leading effect transparency and medium effect transparency. The transparency in the latter case could be the reason of collective interactions of grouped nucleons with the incident particles. 
\end{abstract}
\pacs{25.40.Kv, 25.75.Ag, 25.45.De, 24.10.Jv, 25.40.Ve} 
\maketitle

\section{Introduction} The nuclear transparency (NT) is defined as the ratio of the cross section per target nucleon to that of the one per free nucleon ~\cite{[1]}. It provides a measure of the attenuation effect of the nuclear medium to the propagation of hadrons. Information about the properties of nuclear matter can also be obtained by comparing a given observable in a central nucleus-nucleus collision to the one measured in a pp collision, known as the nuclear modification factor $(R_{AA})$. In the lack of appropriate pp data, which enables one to calculate $R_{AA}$, a ratio of central to peripheral spectra $(R_{cp})$  is used, since ultra-peripheral events look very like elementary collisions. At low energies the parameters given in ~\cite{[1a]} are used to define the NT, which could essentially decrease the cross section of an interaction. Energy dependence of the NT effect gives information about the structure ~\cite{[A]} at low energies, properties ~\cite{[C], [D]} at relativistic energies and phases ~\cite{[E], [F]} of nuclear matter at ultra-relativistic energies. A specific version of the NT is the phenomenon of  color transparency (CT) predicted by quantum chromodynamics (QCD) ~\cite{[2], [3]}. CT explains that hadrons produced in exclusive reactions with high four momentum transfers squared ($Q^{2}$) (with r$_{\perp}$ $\sim$ $(\frac{1}{Q})$) can pass through nuclear matter with reduced interactions ~\cite{[4]}. The increase in the NT indicates the appearance of CT as compared to the predictions of traditional nuclear physics expectations. A lot of efforts have been made to search for the CT effect, including A(p, 2p) ~\cite{[1], [6], [6a], [7], [8]} and A(e, e$^{'}$p) ~\cite{[9], [10], [11], [12], [13], [14]} reactions with protons as a probe for studying the effect.  Mesons were used on the premise that at large $Q^{2}$ the values of r$_{\perp}$ are significantly smaller than the size of the nucleon. The results include, $\rho$-meson production ~\cite{[15], [16]}, diffractive dissociation of pions into di-jets ~\cite{[17]} and pion photoproduction process ~\cite{[18],  [19]}.

	We studied the behavior of some average characteristics (average values of multiplicity, momentum and transverse momentum) of $\pi^{-}$-mesons as a function of $N_{p}$ in an event. Half angle technique ($\theta_{1/2}$)~\cite{[19A], [19B]}  is used to study the nuclear transparency effect. Two different values of the angle ($\theta$=5$^{o}$ and $\theta_{1/2}$=25$^{o}$) are considered for particles produced in p$^{12}$C and d$^{12}$C collision at 4.2A GeV/c. We observed several cases of nuclear transparency effect using the above mentioned method. The two cases are presented below in the results section, as an example of the two effects, which could be the reason of the observed nuclear transparency. In analysis, the experimental results are compared with the Cascade model. 
\section{The Method}
In the previously reported work ~\cite{[19A], [19B]} the half angle technique was applied to extract information on nuclear transparency at high energies. In the first step, using nucleon-nucleon collisions, the half angle was defined which divided the multiplicity of the charged particles produced in the collisions into two equal parts. In the second step, the average values of some characteristic properties of charged particles produced in hadron-nucleus (or nucleus-nucleus) collisions were studied as a function of the baryon density. Those characteristic properties were defined separately for particles with angle smaller than (incone) and greater than (outcone) half angle. The behavior of the average characteristics of the hadrons independent of the baryon density was considered as transparency. It was expected that the incone particles will show transparency due to their high energy and small angle. They can pass the medium with minimum interaction and hence matter will become transparent for those particles. For outcone particles the results are expected to be unlike the former case. They have to lose considerable part of their energy due to interactions, and their characteristics have to be more sensitive to the baryon density. We have followed the method described in ~\cite{[19A], [19B]}.  In this work, we determine the value of $\theta_{1/2}$ equal to 25$^{o}$ using pp collisions at 4.2 GeV/c. $\theta_{1/2}$ divides the particles into the incone and outcone particles. Particles with  $\theta$\textless $\theta_{1/2}$ are the incone particles and those with $\theta$\textgreater $\theta_{1/2}$ are the outcone particles. We described NT as an effect at which the characteristics of  $\pi^{-}$-mesons in proton induced interactions with carbon nuclei (p$^{12}$C) and deuteron induced interactions with carbon nuclei (d$^{12}$C) do not depend on the number of identified protons ($N_{p}$) in an event. $N_{p}$ is used to fix the centrality of collisions because it is connected to the baryon density of nuclear matter. Finally the results are compared with Dubna version of the cascade model~\cite{[20], [21], [23], [24]}. The cascade model is used to describe the general features of a relativistic nucleus-nucleus collision. It is an approach based on simulation, using Monte-Carlo techniques, and is applied to situations where multiple scattering is important. The basic assumptions and procedures of the cascade model are given in~\cite{[20], [21]}. We used the experimental data obtained from the 2-m propane bubble chamber available at the laboratory of high energy physics of the Joint Institute for Nuclear Research (JINR), Dubna, Russia. The chamber was exposed to beams of protons and deuterons, accelerated to a momentum of 4.2A GeV/c, which was kept in a 1.5 Tesla magnetic field at the Dubna Synchrophasotron. A detailed discussion on the interaction mechanism is given in~\cite{[25]}. We, therefore, focused on the negative pions because in our experiment they were identified with highest accuracy amongst all charged particles. We used 12757 p$^{12}$C and 9016 d$^{12}$C interactions in our present work. The methodological details can be found in ~\cite{[26]}. In case of cascade code, we used 50000 p$^{12}$C- and d$^{12}$C-interactions under the same conditions.

\section{Results}
\subsection{Incone $\pi^{-}$-mesons}
The behavior of average multiplicity ($<n>$) of incone $\pi^{-}$-mesons as a function of $N_{p}$ is shown in Fig 1(a). The $<n>$ shows clear transparency  for those $\pi^{-}$-mesons at $\theta$ = 5$^{o}$ denoted by $\star$. For $\theta_{1/2}$ = 25$^{o}$, denoted by $\fullsquare$, the $<n>$ of  $\pi^{-}$-mesons is an increasing function of $N_{p}$. This behavior shows that the slope of the line decreases with decreasing half angle. Results of the cascade model are shown by full line for $\theta_{1/2}$ = 25$^{o}$ and dotted line for $\theta$ = 5$^{o}$ in the graph.
The values of the $<n>$ in d$^{12}$C collision at 5$^{o}$ and 25$^{o}$ as a function of $N_{p}$ are shown in Fig. 1(b). Experimental results are given by the geometrical symbols whereas the cascade results are shown by lines. The results in d$^{12}$C -interactions are similar to that of p$^{12}$C. The $<n>$ in d$^{12}$C as a function of the $N_{p}$ shows positive slope at 25$^{o}$ and transparency at 5$^{o}$. The above observations are verified by fitting the two graphs using linear function $<n>$ = A+B*$N_{p}$.

\begin{figure}[t]
\centering
\includegraphics[width=2.5in]{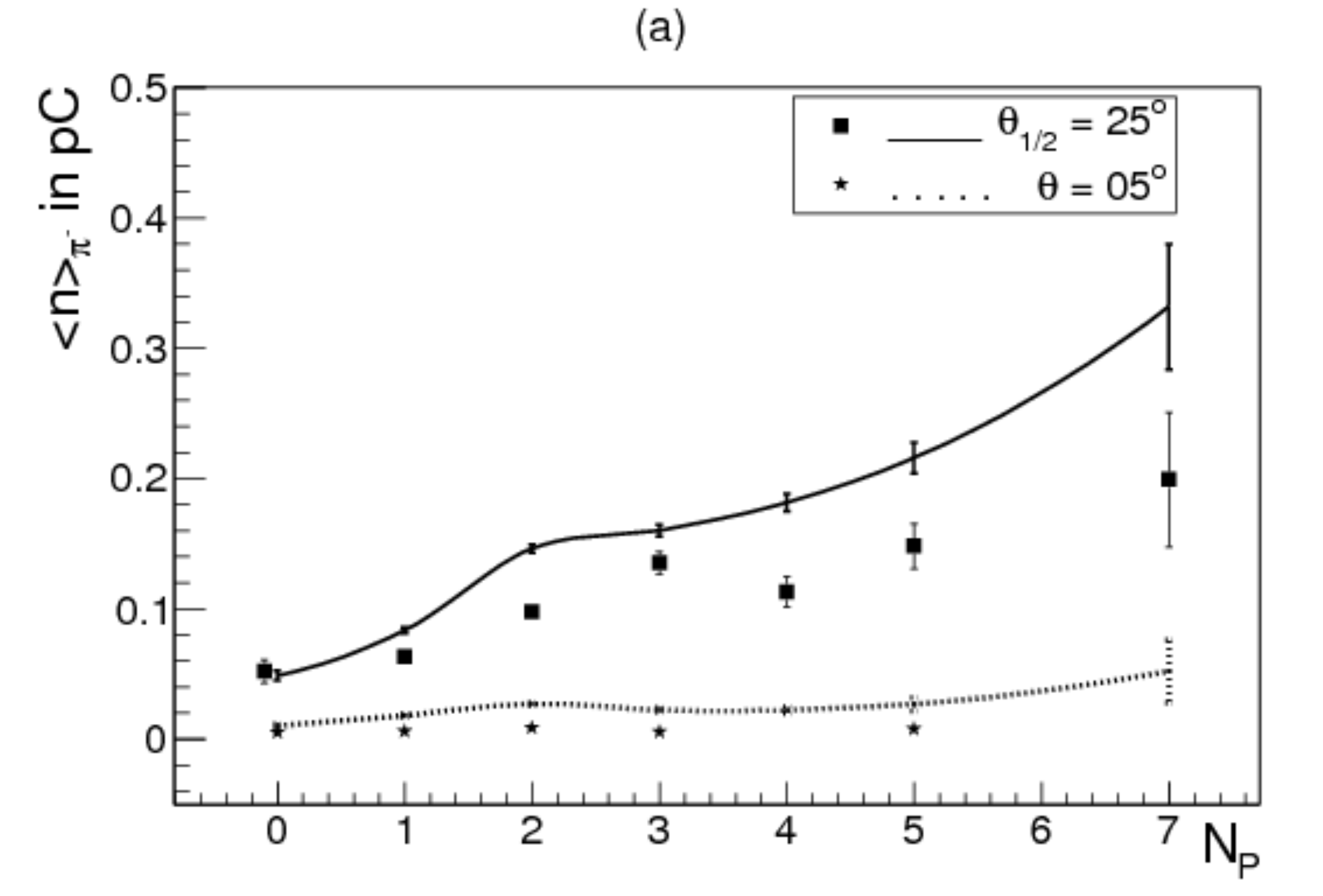}
\includegraphics[width=2.5in]{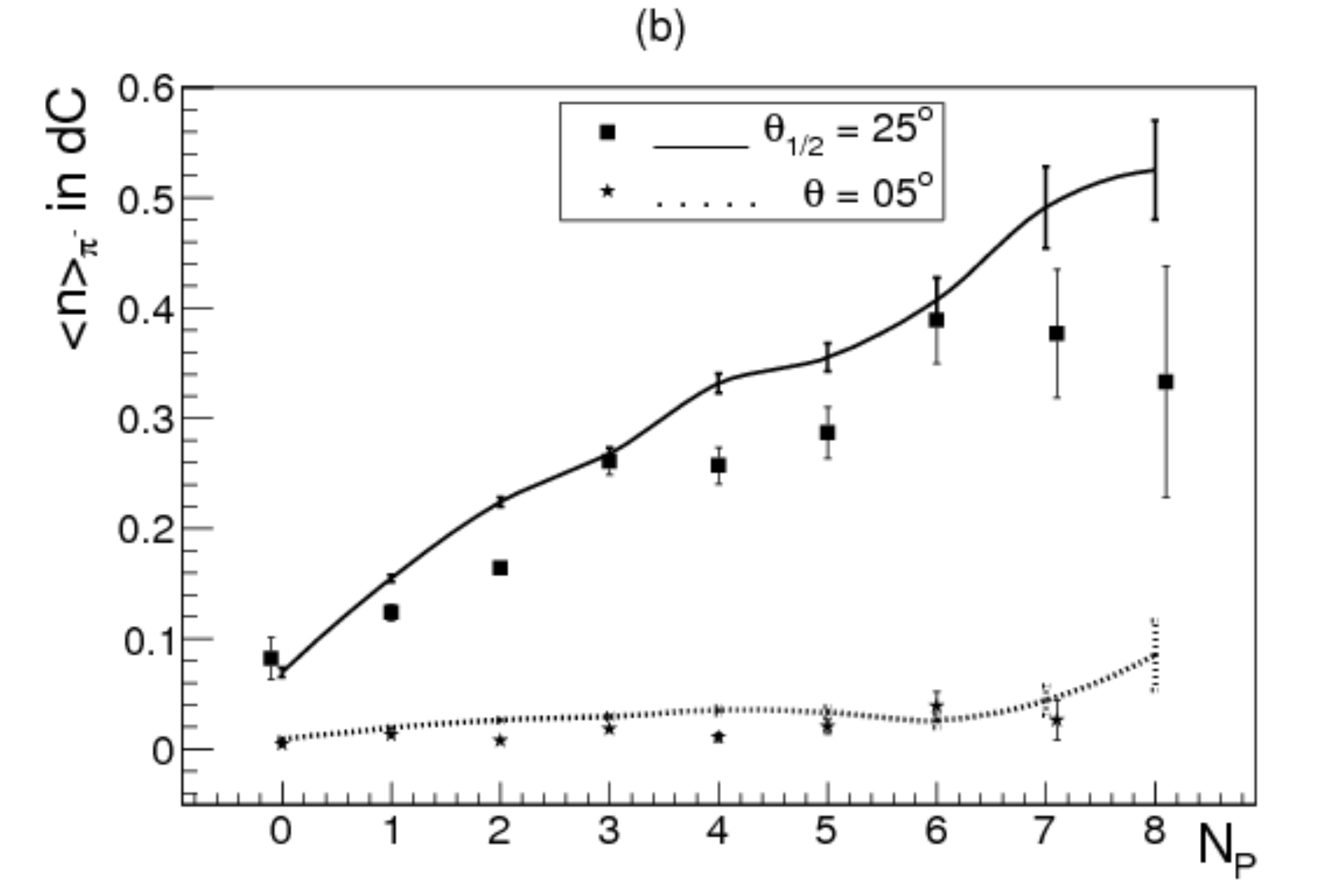}
\includegraphics[width=2.5in]{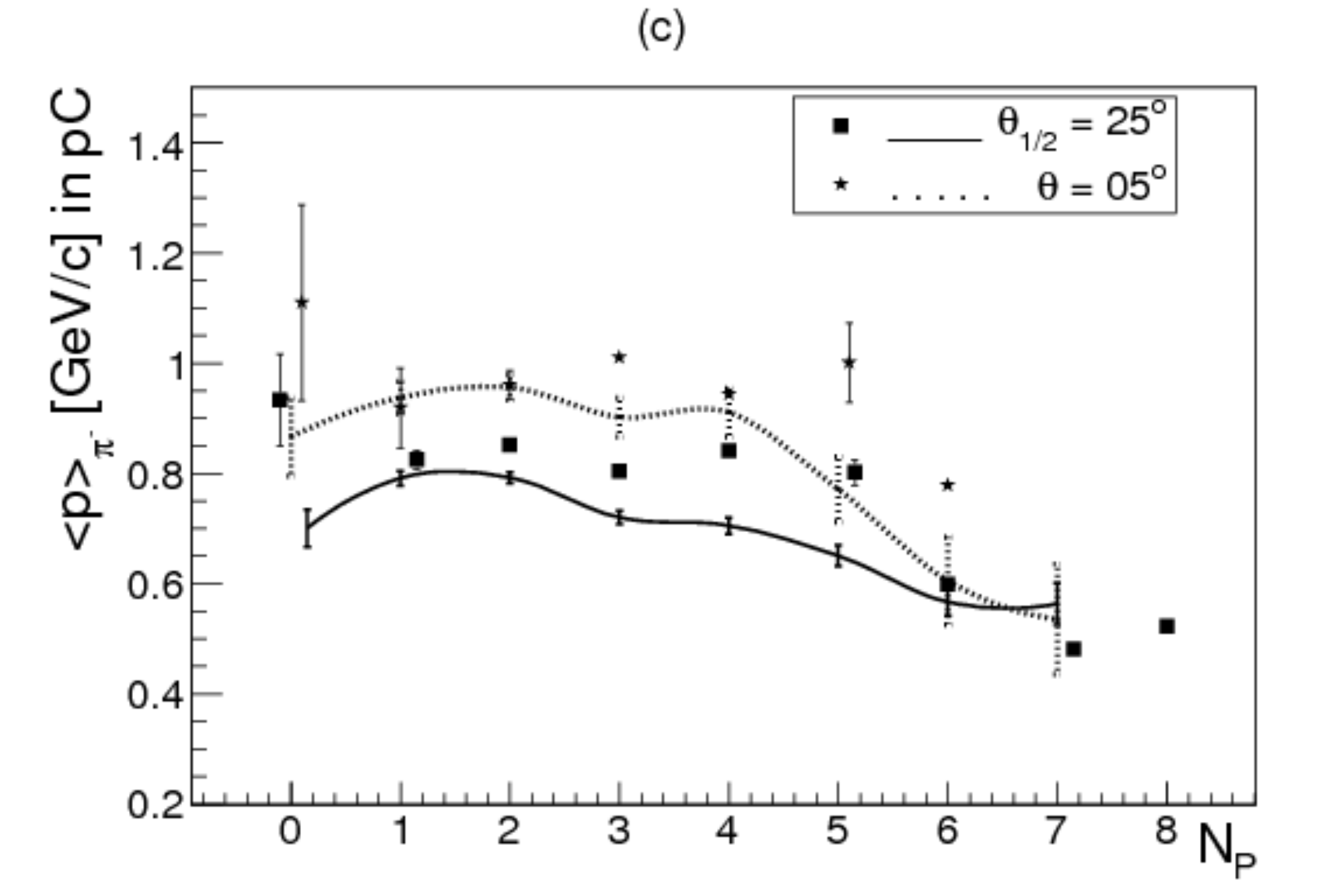}
\includegraphics[width=2.5in]{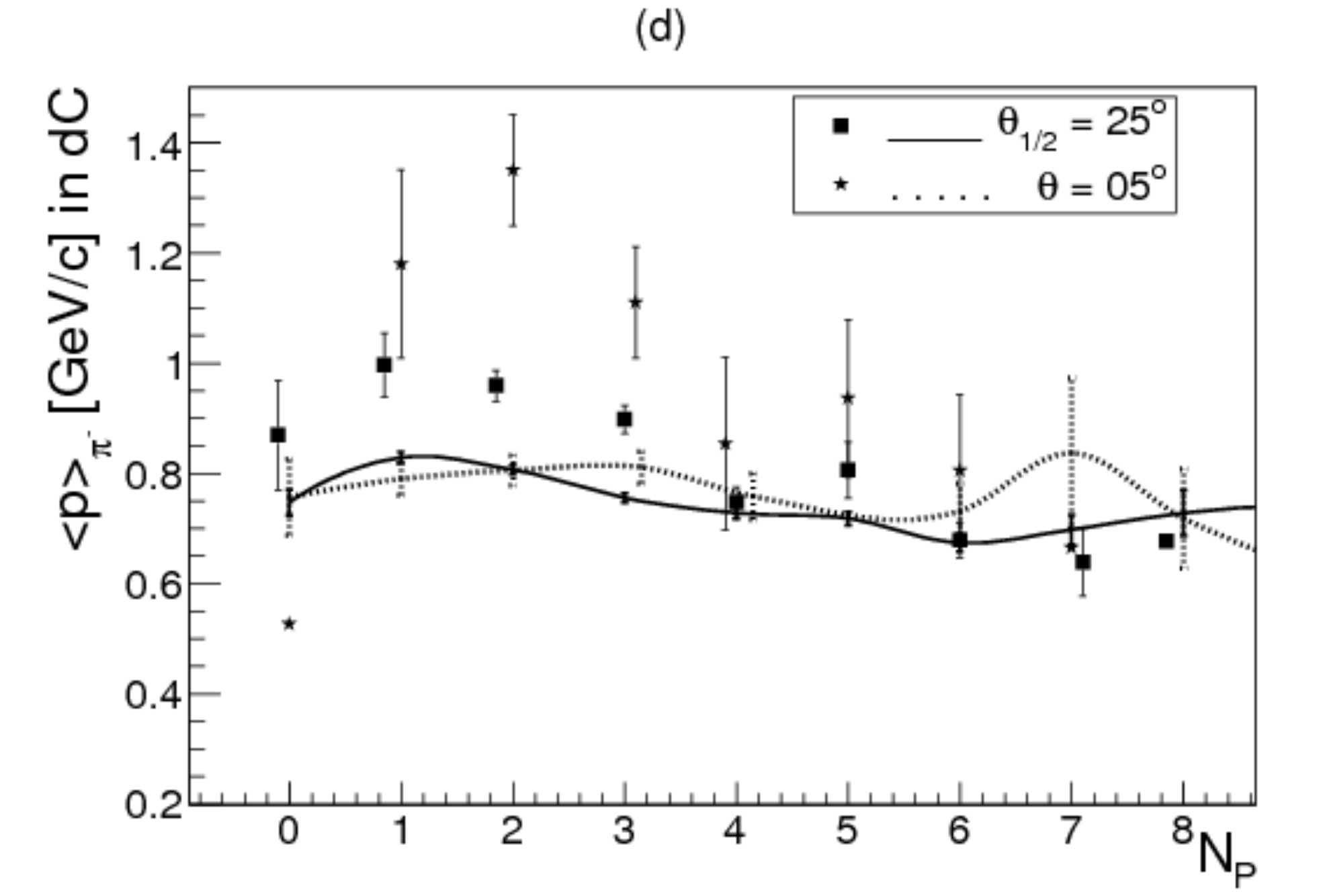}
\includegraphics[width=2.5in]{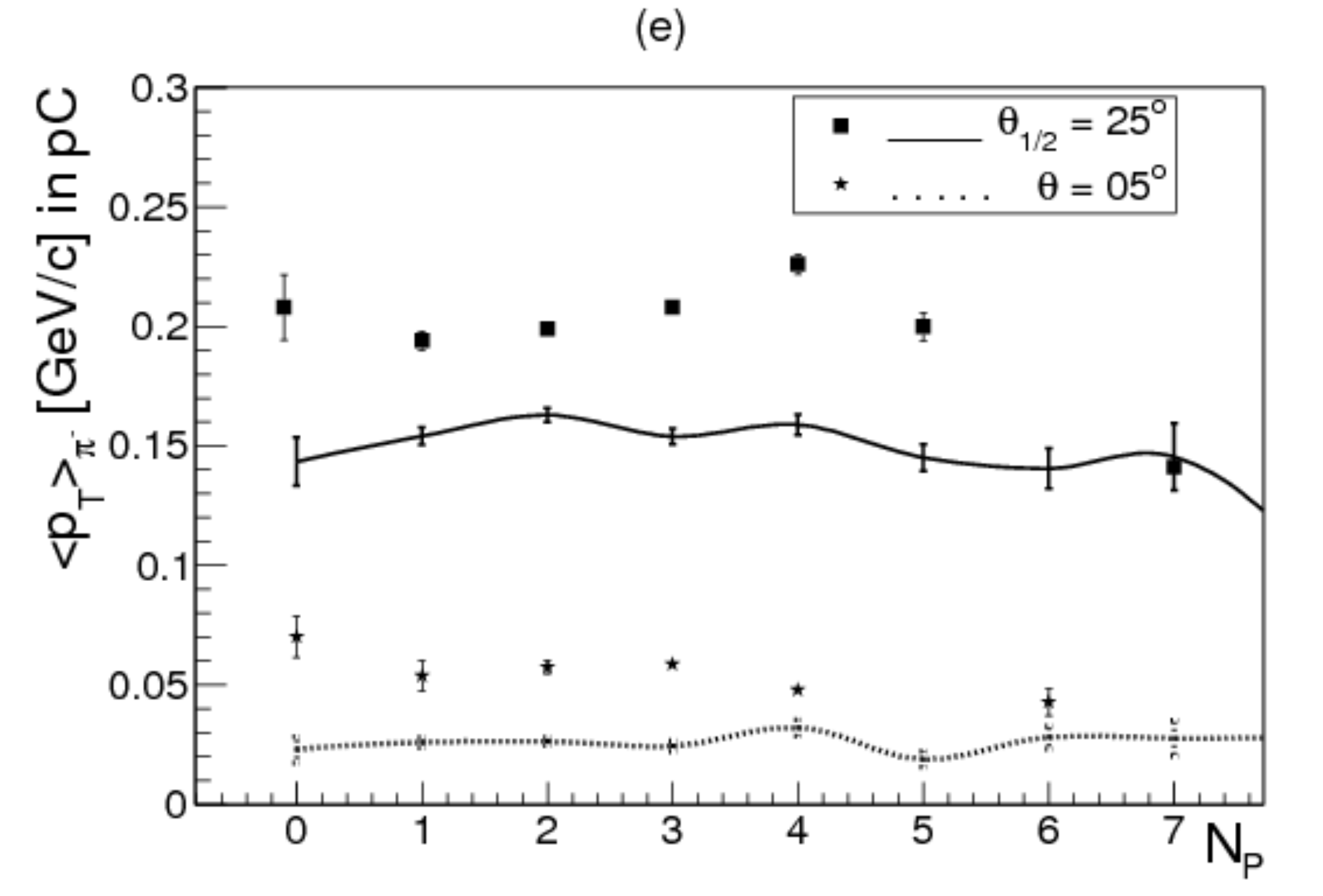}
\includegraphics[width=2.5in]{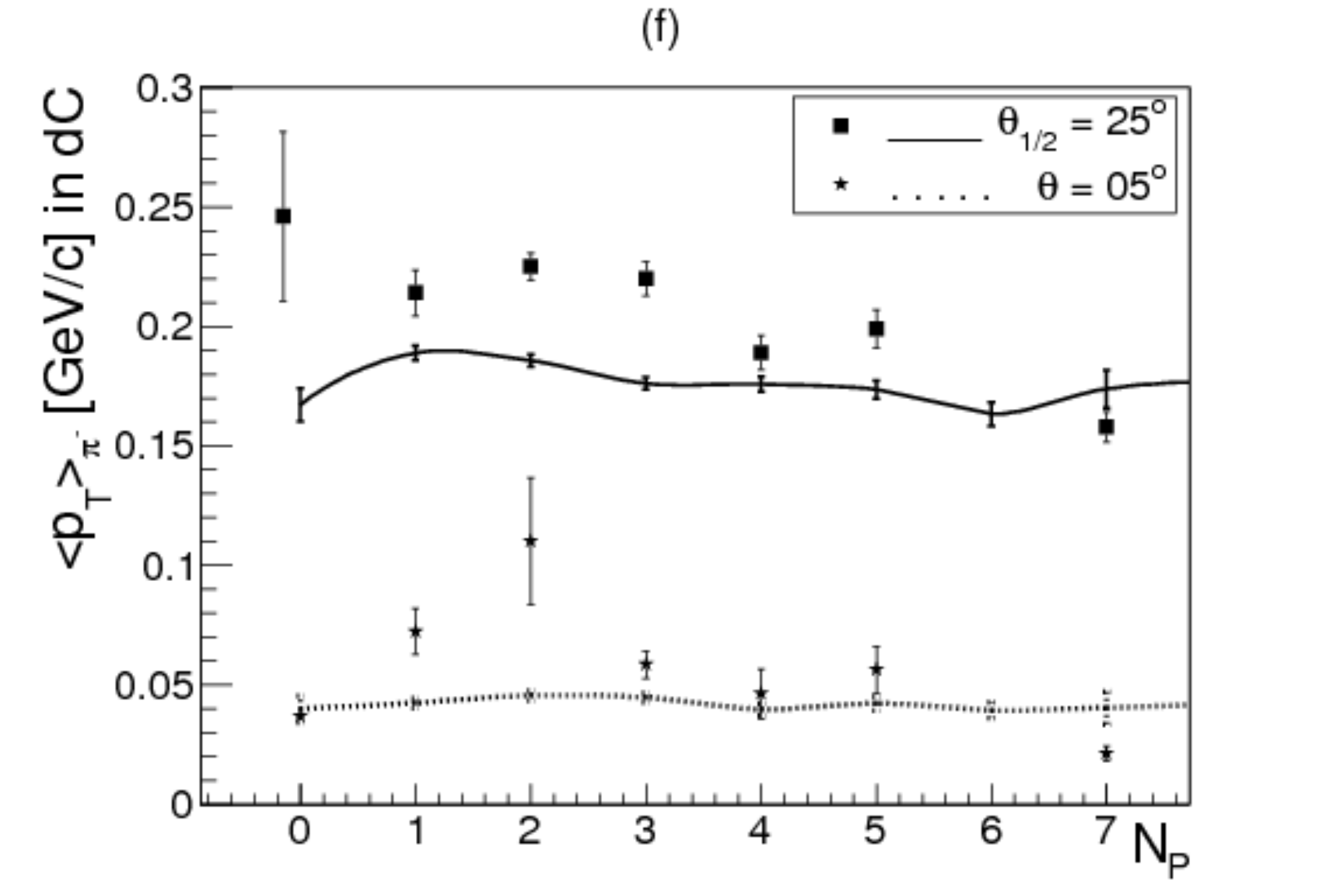}
\caption{Behavior of the average characteristics of incone $\pi^{-}$-mesons as a function of the number of identified protons ($N_{p}$). Fig 1(a) is the behavior of average multiplicity ($<n>$) of incone $\pi^{-}$-mesons as a function of $N_{p}$  in proton induced interaction with carbon nuclei (p$^{12}$C), 1(b) is the $<n>$ of incone $\pi^{-}$-mesons as a function of $N_{p}$ in d$^{12}$C- interactions, 1(c) is the $<p>$ versus $N_{p}$ in p$^{12}$C and 1(d) is d$^{12}$C- interactions, 1(e) is the $<p_{T}>$ versus $N_{p}$ in p$^{12}$C and 1(f) is the $<p_{T}>$ versus $N_{p}$ in d$^{12}$C- interactions.}
\label{fig1}
\end{figure}
Comparison of the two results demonstrates that the transparency is observed only for the  $\pi^{-}$-mesons which lies in the smallest angle. 
The values of incone $\pi^{-}$-meson's average momentum $<p>$, in proton induced carbon interaction, as a function of the $N_{p}$ for the experimental and code data are given in Fig. 1(c) and 1(d). There are two regions for the behavior of the $<p>$ of those $\pi^{-}$-mesons. In the first region ($N_{p}$=0-5) the values of $<p>$ decreases slowly with $N_{p}$ and in the second one (for the values of $N_{p}$\textgreater 5) the values of $<p>$ decrease sharply. Those fast and small-angle incone meson's transparency could be connected to the leading nucleons ~\cite{[D]}. They could appear as a result of charge exchange reactions ~\cite{[28]} of the leading nucleon N+N$\rightarrow$N+N+ $\pi$. The $\pi^{-}$ -mesons, produced by the leading nucleons, will possess high energies and small angles. The idea is confirmed by the comparison of p$^{12}$C, with d$^{12}$C data given in Fig 1(d). The values of $<$p$>$ in case of experimental data are large compared to that in the cascade model and are even higher than the same values obtained for p$^{12}$C.  In p$^{12}$C the difference between model and experimental data is smaller than that of the d$^{12}$C-data, because deuteron is a two nucleon system with one extra projectile neutron. This additional leading neutron could be an extra source of producing the fast and low angle $\pi^{-}$ -mesons. Leading particles are highly energetic particles in an event which could give-up a part of their energy during interaction.  The particles will have maximum energy in an event and would be identified in an experiment as incone particles due to their high energy and low angle. With high energy they are able to pass through the medium without losing a large fraction of their initial energy, making the medium transparent. Since we have observed the transparency for those $\pi^{-}$-mesons lies in the smallest angle, therefore, they are the highest energy particles. These high energy small angle $\pi^{-}$-mesons could be produced by high energy nucleons during their charge exchange interactions. Most of the kinetic energy during the charge exchange interaction will be taken by the $\pi^{-}$-mesons, because the lighter particles will carry higher kinetic energies. Finally the behavior of average transverse momentum ($<p_{T}>$) of incone $\pi^{-}$-mesons as a function of $N_{p}$ are shown in Fig 1(e) and 1(f) for p$^{12}$C and d$^{12}$C data respectively. The behavior in these two cases also show that the $<p_{T}>$ of the mesons is a slowly decreasing function of $N_{p}$. Fitting the data in a linear function also shows a negative slope. The leading nucleons during their passage through medium transfer only a small fraction of their initial energy. That is why these incone $\pi^{-}$-mesons show transparency for the $<n>$ but a slight decrease has occurred in $<p>$ and $<p_{T}>$.
\subsection{Outcone $\pi^{-}$-mesons}
The average multiplicity $<n>$ of the outcone $\pi^{-}$-mesons in p$^{12}$C- interactions at  $\theta$ = 5$^{o}$ and $\theta_{1/2}$ = 25$^{o}$ as a function of the $N_{p}$ are shown in Fig.2(a). The results from the d$^{12}$C data under the same conditions are given in Fig. 2(b). The values of $<n>$ in p$^{12}$C and d$^{12}$C data increase linearly with $N_{p}$ in both cases. The linear fitting of the data results in a steeper slope in cascade model as compared to the slope in experimental data. Comparison of the slopes shows that the slope in case of cascade model has substantial increase (e.g for 5$^{o}$ from 0.21-0.32) in d$^{12}$C as compared to p$^{12}$C results, while there is a very little increase (e.g for 5$^{o}$ from 0.16-0.18) in the slope of experimental data in d$^{12}$C. Thus, one can say that cascade model could not describe $N_{p}$ –dependence for outcone $\pi^{-}$-mesons $<n>$. Consequently, one can suppose that there might be some mechanisms that could have reduced the increasing multiplicity of the $\pi^{-}$-mesons in the outcone. A possible mechanism could be the collective interactions of grouped nucleon in the target with projectile particle. So in case of p$^{12}$C interaction the particle-tube interaction is responsible to prevent the increasing multiplicity of these mesons, while in case of d$^{12}$C the tube-tube does the same job. That is why the multiplicity in the latter case remained almost the same.
\begin{figure}[t]
\centering
\includegraphics[width=2.5in]{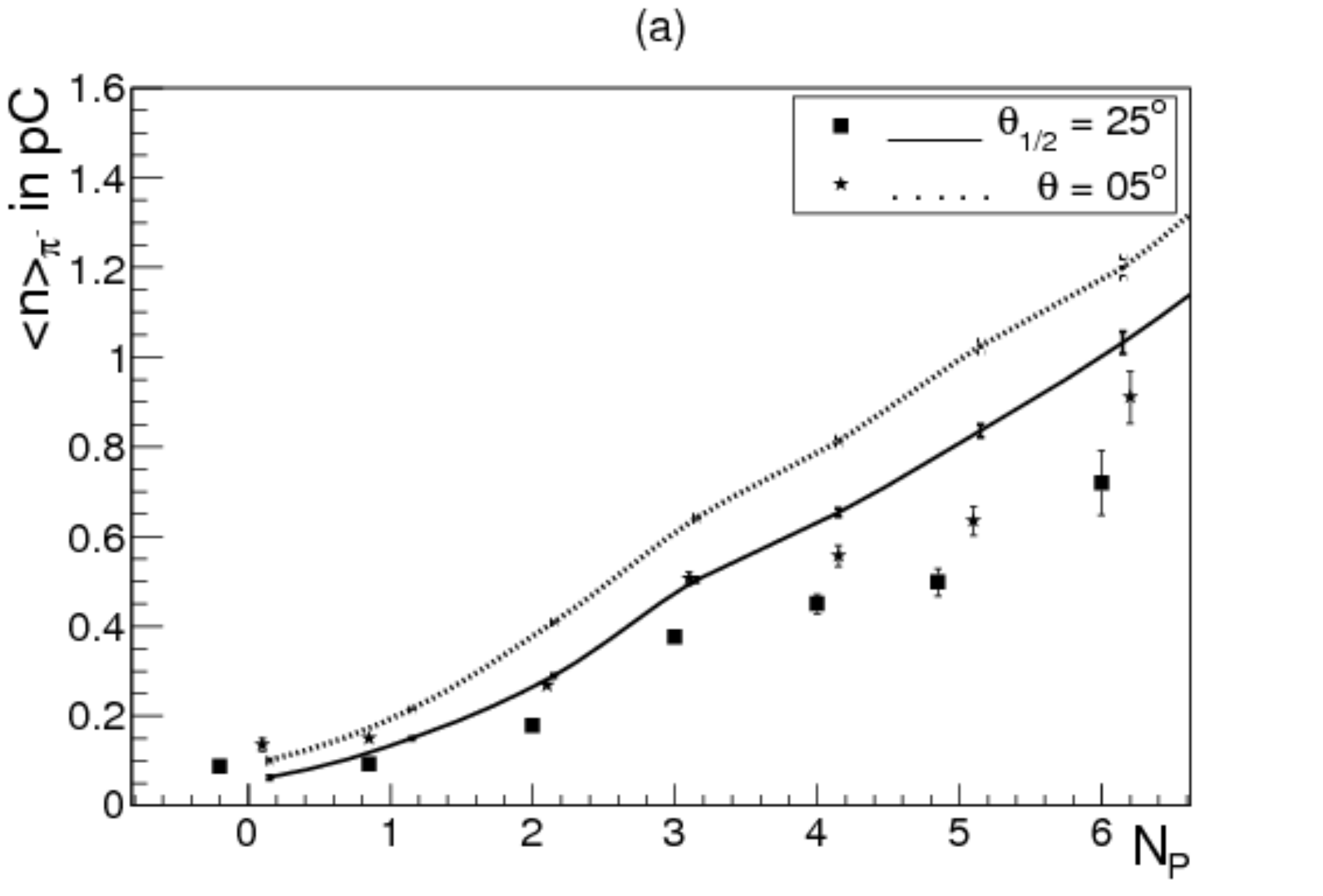}
\includegraphics[width=2.5in]{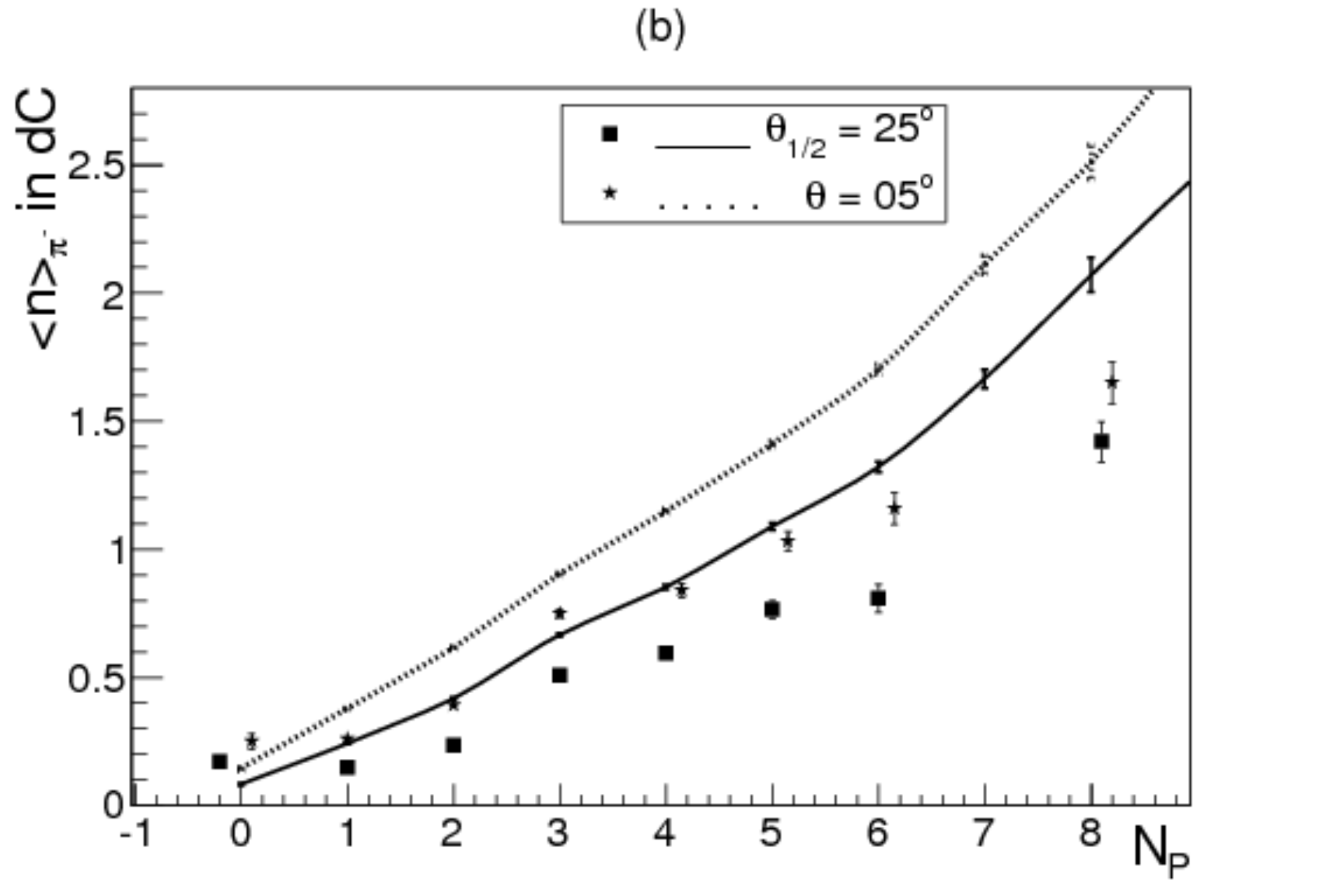}
\includegraphics[width=2.5in]{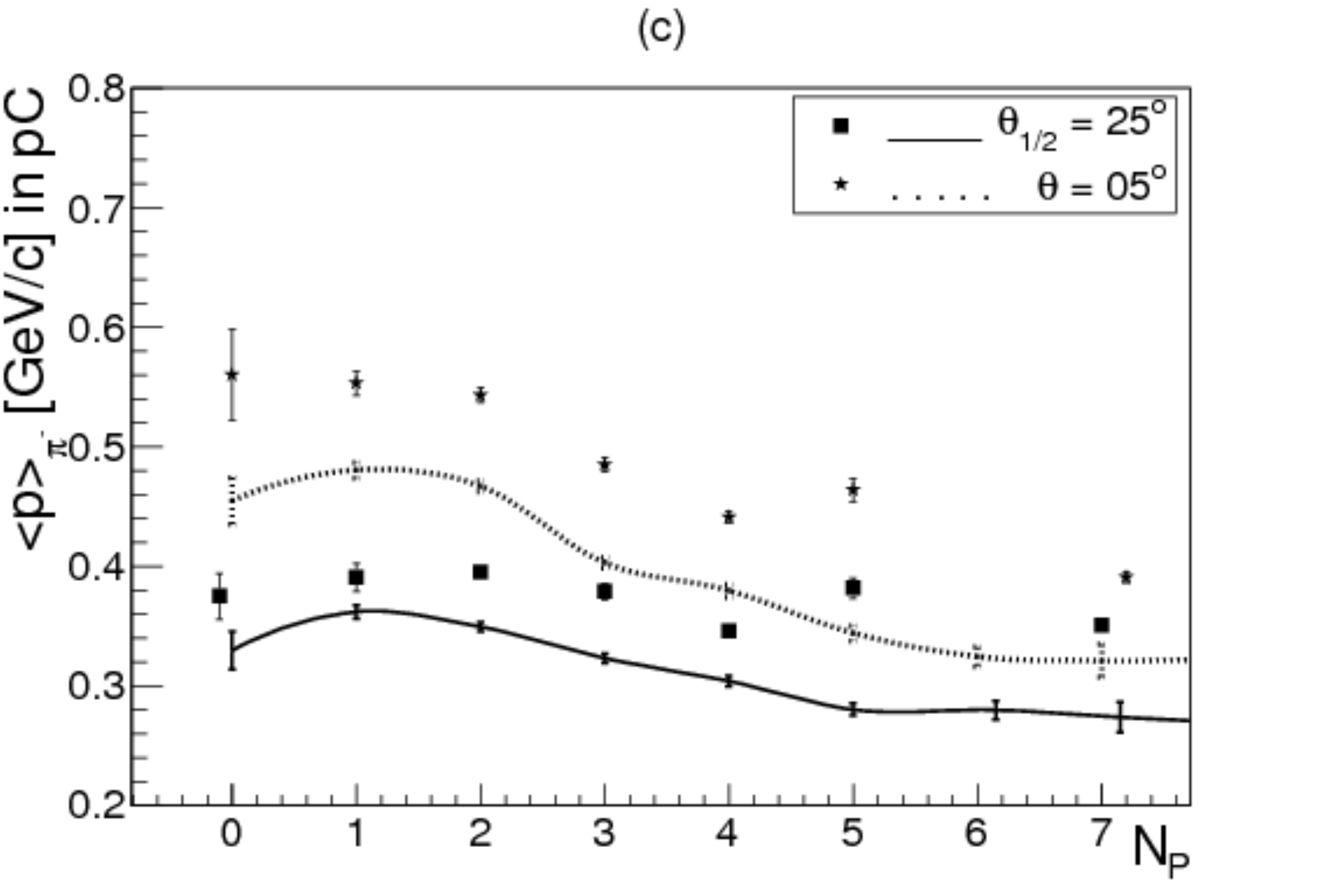}
\includegraphics[width=2.5in]{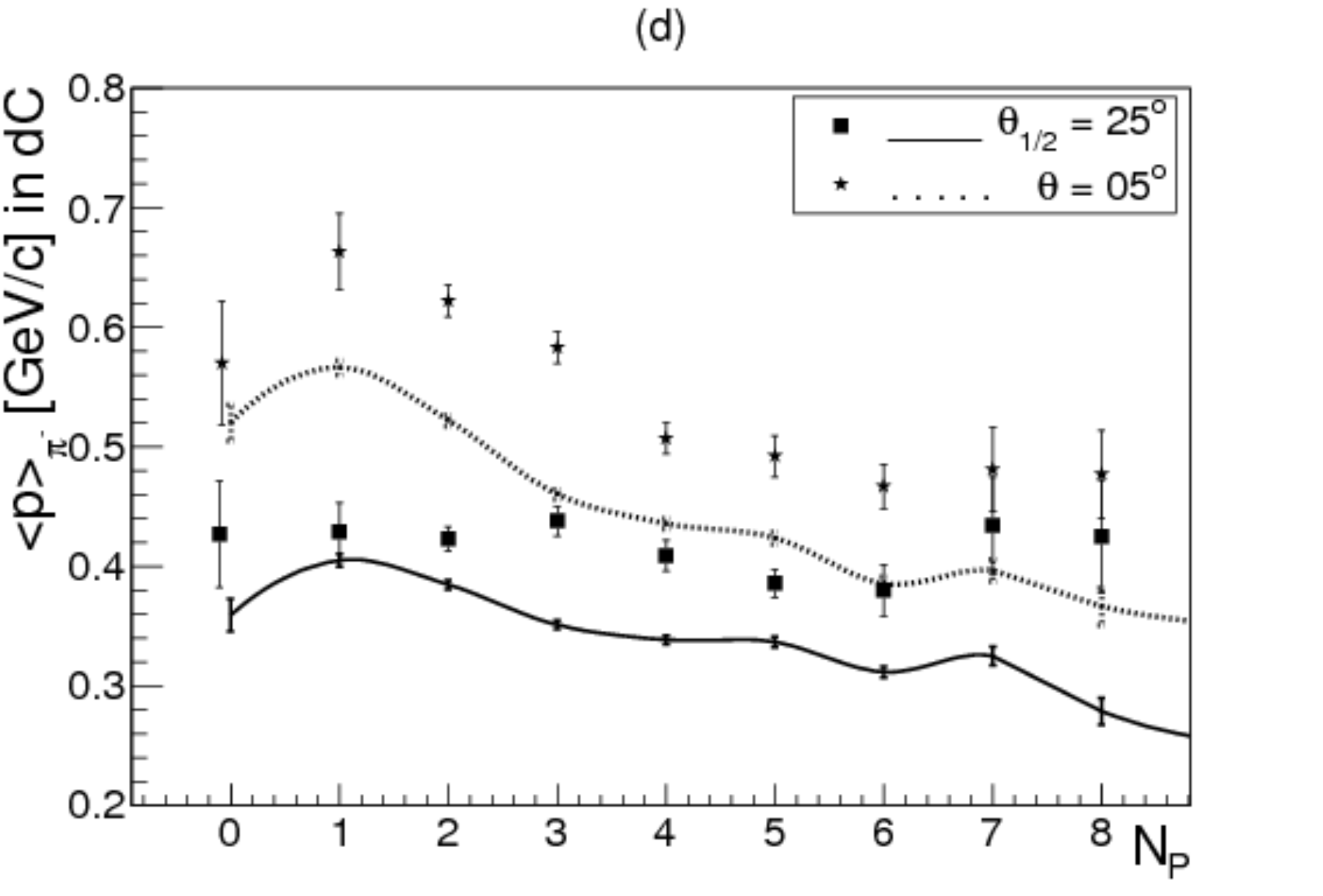}
\includegraphics[width=2.5in]{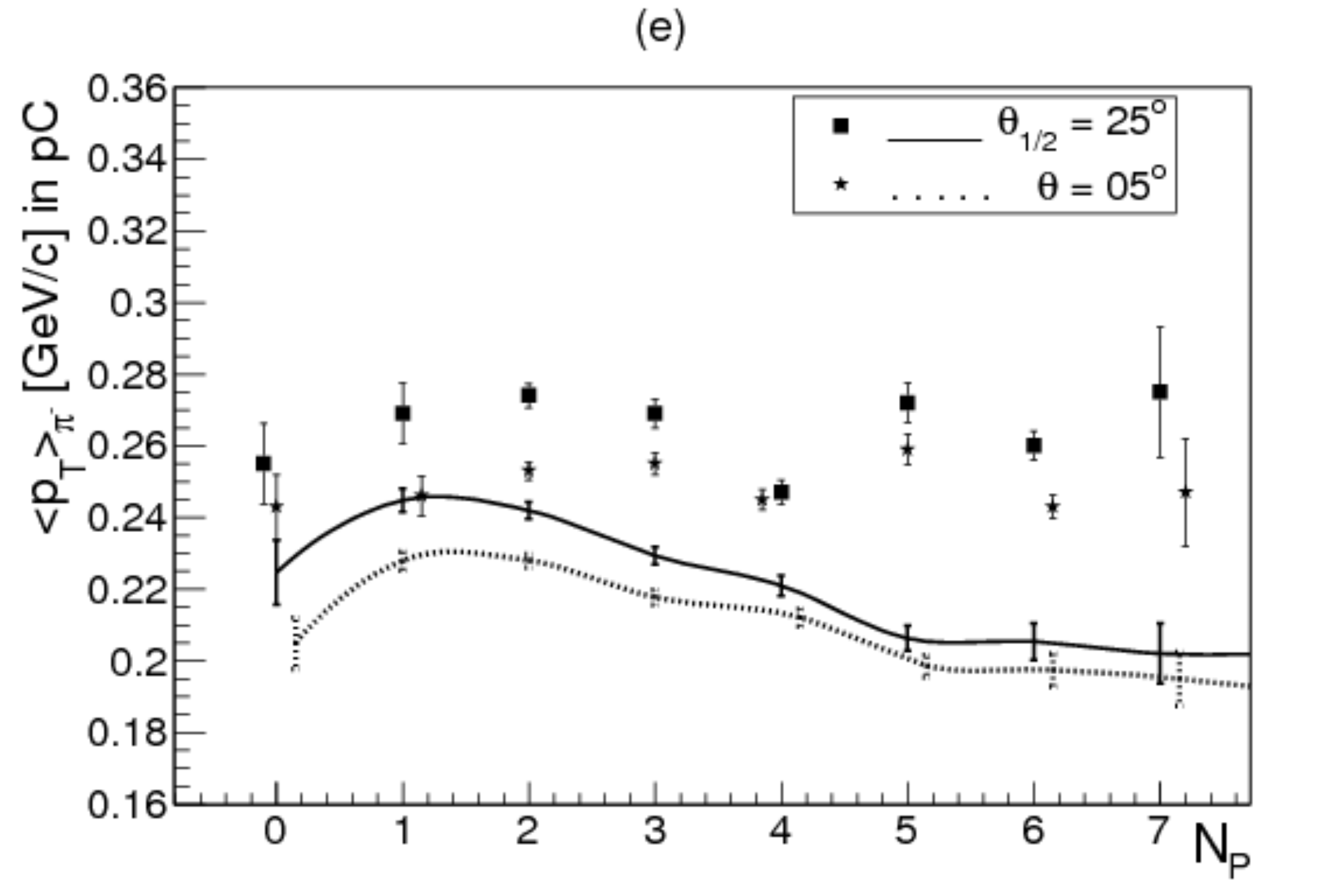}
\includegraphics[width=2.5in]{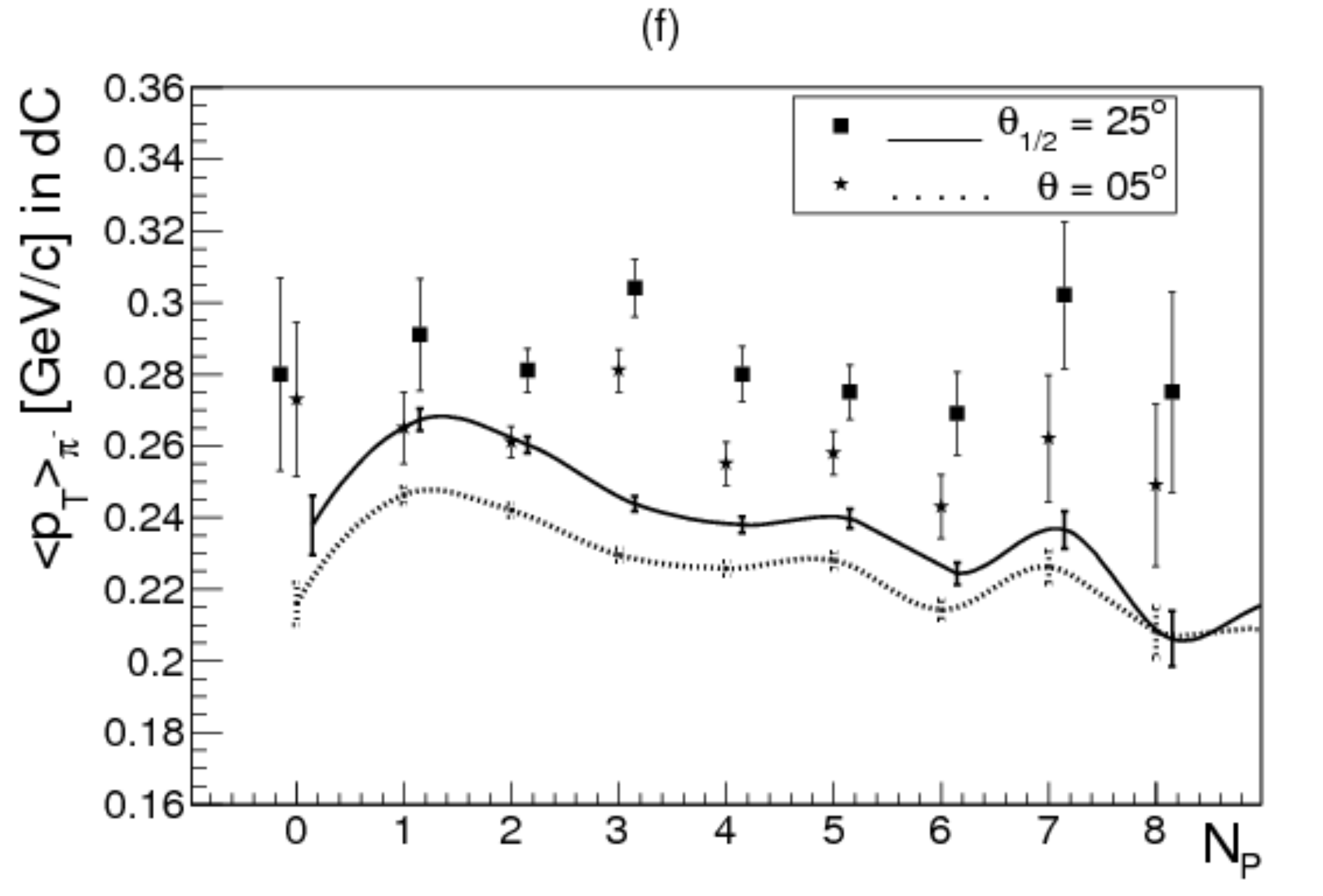}
\caption{Behavior of the average characteristics of the outcone $\pi^{-}$-mesons as a function of number of identified protons ($N_{p}$). Fig 2(a) is the behavior of average multiplicity ($<n>$) of outcone $\pi^{-}$-mesons as a function of $N_{p}$  in proton induced interaction with carbon nuclei (p$^{12}$C), 2(b) is the $<n>$ of outcone $\pi^{-}$-mesons as a function of $N_{p}$ in d$^{12}$C- interactions, 2(c) is the $<p>$ versus $N_{p}$ in p$^{12}$C and 2(d) is d$^{12}$C- interactions, 2(e) is the $<p_{T}>$ versus $N_{p}$ in p$^{12}$C and 2(f) is the $<p_{T}>$ versus $N_{p}$ in d$^{12}$C- interactions.}
\label{fig2}
\end{figure}
The values for the $<$p$>$ as a function of the $N_{p}$ for these pions in p$^{12}$C and d$^{12}$C data are given in Fig. 2(c) and 2(d) respectively for the two angles as shown. The results of the experimental and the code data are different. At $\theta_{1/2}$ = 25$^{o}$ the experimental data shows transparency whereas the cascade results show a decreasing behavior with increasing $N_{p}$.  Data obtained from the code cannot explain the transparency. The code data demonstrates the existence of two regions for the behavior of the $<$p$>$ in p$^{12}$C. In the first region the values decrease rapidly as compared to a slow decrease in the second region with $N_{p}$. The $<$p$>$ at $\theta$ = 5$^{o}$ is a decreasing function of $N_{p}$ in both cases. In case of experimental data, the higher values of $<p>$ at $\theta$ = 5$^{o}$ around $N_{P}$ = 1 of these outcone meson's indicate the presence of the fast leading meson's in this category. The results of the d$^{12}$C are the same as that of p$^{12}$C.
Fig. 2(e) and 2(f) demonstrate $<p_{T}>$ as a function of the $N_{p}$ for the p$^{12}$C and d$^{12}$C data respectively. A clear difference can be observed between experimental and code data. Experimental data shows transparency in both p$^{12}$C and d$^{12}$C data.  The code data consists of two regions. In the first region the value of $<p_{T}>$ decrease rapidly, whereas in the second region it decrease slowly. 
Thus, we could say that the experimental data of the out cone $\pi^{-}$-mesons demonstrate some transparency which is not described by the cascade model. This behavior may not be the reason of the leading effect due to the fact that the out cone  $\pi^{-}$-mesons are secondary produced particles with small energy and large angle. In both cases transparency seems to be more clear at level of $<p_{T}>$ $\approx$ 0.28 GeV/c. The size of pion radiation region $(R)$ could be approximately defined from the  pions $<p_{T}>$  as  R $\sim$ $\frac{1}{<p_{T}>}$. Using this expression the value of  $R$ obtained is equal to 3.6 fm. Diameter of the carbon nucleus is 3.67 fm, which is the same as that of the outcone pions radiation region. We can see that the projectile did not change the values of the $R$, because in p$^{12}$C and d$^{12}$C interactions the values of $R$ = 3.6 fm is constant. Thus, we can say that a reason for the observed transparency is that the size of pion radiation region is not affected during the interaction with increasing the baryon density as well as with the mass of the projectile.
A different mechanism is responsible for the transparency observed in this process, which seems to be connected to some particular properties of the medium. The various models, that are used so far, for the high energy particle-nucleus and nucleus-nucleus interactions, are divided into two categories~\cite{[29]}. We are concerned only with the second category of models that includes all models which assume that particle-nucleus collisions is a single step process, where a few nucleons in the nucleus, interact collectively with the incident particle. In such models the effective center of mass energy available is approximately given by 
\begin{center}
$s_{eff}$ $\approx$ 2$m_{eff}P_{lab}$
\end{center}
where $m_{eff}$ is the mass of the system that interacts collectively with the incident particle. Its value is of the order of a few times the mass of a single nucleon. $P_{lab}$ is the momentum of the incident particles. We think that the observed transparency could be the results of the collective interaction of grouped nucleons with the proton in p$^{12}$C data and with grouped nucleon (deuteron) in d$^{12}$C data. Using the following expression from Coherent Tube Model (CTM) ~\cite{[29]} for the average multiplicity of secondary charged particles ($<n(s)>$) as a function of the number of fast protons ($N_{p}$), we can get roughly the number of protons in tube ($i_{p}$) which responds collectively to the incident particles. In the CMT ~\cite{[29]} the multiplicity of secondary charged particles is given by the expression: 
\begin{center}
                                             $<n(s)>_{ip}$ $\approx$ $<n(s)> _{pp}$ x $(< i_{p}> )^{\alpha}$
\end{center}
where   $<n(s)>_{ip}$ and $<n(s)> _{pp}$  are the  multiplicities of the secondary charged particles produced in tube with $i_{p}$ nucleons and in pp-interactions respectively at same energy in c.m.  A and Z are the mass number and charge number of target nucleus respectively (for carbon nucleus A = 12, Z = 6). For estimation we take $\alpha$ $\sim$ $\frac{1}{4}$ ~\cite{[30]} which gives $i_{p}$ = 3.1 $\pm$ 0.3 which does not depend on “half angle”.
 The diameter of the carbon nucleus is about 3.6 fm which can accommodate $\sim$ 2.5 nucleons placed side by side to each other. Our calculated value for the number of nucleons in the tube is close to the number of nucleons in the diameter of the carbon nucleus. This result gives us an opportunity to say that the collective interactions of grouped nucleons in the nucleus is a reason for the observed transparency in outcone $\pi^{-}$-mesons.
\section{Summary and Conclusion}
In search of the nuclear transparency effect for particular properties of nuclear matter, we studied the behavior of the average multiplicity, momentum and transverse momentum of $\pi^{-}$-mesons as a function of a number of identified protons in an event using p$^{12}$C- and d$^{12}$C- interactions at 4.2A GeV/c. We used half angle technique which divides the particle multiplicity in the nucleon-nucleon collisions at the same energy into two equal parts; incone particles with polar angle smaller than the half angle; outcone particles with polar angle greater than the half angle. In our investigations we observed some signals of the nuclear transparency effect. We found that the average values of 

-	multiplicity, momentum and transverse momentum of incone $\pi^{-}$-mesons, and 

-	momentum and transverse momentum of outcone $\pi^{-}$-mesons';\\
do not depend completely or partially in some cases, on the number of identified protons –some signals on appearance of the nuclear transparency effect.

-	Comparison of the experimental results with those obtained from the Dubna Cascade Model indicates that the signal on transparency for the incone $\pi^{-}$-mesons is due to the leading particles. Because these are the $\pi^{-}$-mesons which have high energy, small polar angle and their investigated characteristics are described satisfactorily by the model, they could be produced during charge exchange interactions of the leading nucleons. 

-	the signal on transparency coming from the behavior of the average characteristics of the outcone $\pi^{-}$-mesons could not be explained on the basis of the leading effect, because the effect is observed for the outcone $\pi^{-}$-mesons, having low energy and large angle. Furthermore, the model also could not describe the results satisfactorily.  

-	using the values for the average transverse momentum for outcone $\pi^{-}$-mesons, we defined roughly  that the size of radiation region is R $\sim$ 3.6 fm;

-	simple estimation of the number of nucleons from the average values of multiplicity of the pions  using the Coherent Tube Model  gives the values $\sim$ 3.1;

-	the last two points could be considered as some evidence, that the collective nucleons in the coherent tube is the reason of the transparency effect. \\
\\


\begin{thebibliography}{100}
\bibitem{[1]} A. S. Carroll et al., \emph{"Nuclear Transparency to Large-Angle pp Elastic Scattering"}, Phys. Rev. Lett. 61, (1988) 1698.
\bibitem{[1a]} H. A. Bethe, \emph{"A Continuum Theory of the Compund Nucleus"}, Phys. Rev. 57, (1940) 1125.
\bibitem{[A]} S. Szilner, F. Haas and Z. Basrak, "Weak absorption and resonances in light heavy ion reactions induced by the non-alpha-type 14C nucleus", FIZIKA B 12, 2, (2003) 117.
\bibitem{[C]} H. Satz. "Deconfinement and percolation.", Nucl. Phys. A, 642, (1998), 130.
\bibitem{[D]} M. Ajaz, M. K. Suleymanov, K. H. Khan and A. Zaman, "Searching for the properties of nuclear matter using proton-carbon and deuteron-carbon collisions at 4.2A GeV/c. " Int.J.Mod.Phys. E, 21(12) (2012) 1250095
\bibitem{[E]} W. Cosyn et al., "Color transparency and short-range correlations in exclusive pion photo- and electroproduction from nuclei.", Phys. Rev., C 77, (2008) 034602.
\bibitem{[F]} M. Ajaz, M. K. Suleymanov, K. H. Khan and A. Zaman, "Study of the behavior of the nuclear modification factor in freeze-out state", Chin.Phys. C 37, (2013) 024101.
\bibitem{[2]} S. J. Brodsky in: E.W. Kittel, W. Metzger, A. Stergiou (Eds.), Proceedings of the XIII International Symposium on Multiparticle Dynamics, Volendam, The Netherlands, World Scientific, Singapore, 1982, p. 963.
\bibitem{[3]} A. H. Mueller in: J. Tran Thanh Van (Ed.), Proceedings of the XVII Rencontre de Moriond, Les Arcs, France, Editions Frontieres, Gif-sur-Yvette, 1982, p. 13.
\bibitem{[4]} S. J. Brodsky and A. H. Mueller, \emph{"Using nuclei to probe hadronization in QCD"}, Phys. Lett. B 206 (1988) 685.
\bibitem{[6]} I. Mardor et al., \emph{"Effect of multiple scattering on the measurement of nuclear transparency"}, Phys. Rev. C, 46, (1992) 761.
\bibitem{[6a]} I. Mardor et al., \emph{"Nuclear Transparency in Large Momentum Transfer Quasielastic Scattering"}, Phys. Rev. Lett. 81, (1998) 5085.
\bibitem{[7]} A. Leksanov et al., \emph{"Energy Dependence of Nuclear Transparency in $^{12}$C(p, 2p) Scattering"}, Phys. Rev. Lett. 87, (2001) 212301.
\bibitem{[8]} J. L. S. Aclander et al., \emph{"Nuclear transparency in 90$^{o}_{c.m}$ quasielastic A(p,2p) reactions"}, Phys. Rev. C 70, (2004) 015208.
\bibitem{[9]} G. Garino et al., \emph{"Proton propagation in nuclei studied in the (e, e$^{'}$p) reaction"}, Phys. Rev. C 45, (1992) 780.
\bibitem{[10]} N. C. R. Makins et al., \emph{"Momentum Transfer Dependence of Nuclear Transparency from the Quasielastic $^{12}$C(e, e$^{'}$p) Reaction"}, Phys. Rev. Lett. 72, (1994) 1986.
\bibitem{[11]} T. G. O’Neill et al., \emph{"A-dependence of nuclear transparency in quasielastic A(e, e$^{'}$p) at high Q$^{2}$"}, Phys. Lett. B 351, (1995) 87.
\bibitem{[12]} D. J. Abbott et al., \emph{"Quasifree (e,e$^{'}$p) reactions and proton propagation in nuclei," }, Phys. Rev. Lett. 80, (1998) 5072.
\bibitem{[13]} K. Garrow et al., \emph{"Nuclear transparency from quasielastic A(e, e$^{'}$p) reactions up to Q$^{2}$ = 8.1 (GeV/c)$^{2}$"}, Phys. Rev. C 66, (2002) 044613.
\bibitem{[14]} D. Dutta et al. (Jefferson Lab E91013), \emph{"A Study of the Quasi-elastic (e ,e$^{'}$p) Reaction on $^{12}$C, $^{56}$Fe, and $^{97}$Au"}, Phys. Rev. C 68, (2003) 064603.
\bibitem{[15]} M. R. Adams et al. (E665), \emph{“Measurement of Nuclear Transparencies from Exclusive $\rho^{o}$ Meson Production in Muon-Nucleus Scattering at 470 GeV”}, Phys. Rev. Lett. 74, (1995) 1525.
\bibitem{[16]} A. Airapetian et al. (HERMES), \emph{"The $Q^{2}$ dependence of nuclear transparency for exclusive $\rho^{o}$ production"}, Phys. Rev. Lett. 90, (2003) 052501.
\bibitem{[17]} E. M. Aitala et al. (E791), \emph{"Observation of color-transparency in diffractive dissociation of pions"}, Phys. Rev. Lett. 86, (2001) 4773.
\bibitem{[18]} D. Dutta et al. (Jefferson Lab E940104), \emph{"Nuclear transparency with the $\gamma$n $\rightarrow$ $\pi^{-}$p process in $^{4}$He"}, Phys. Rev. C 68, (2003) 021001(R).
\bibitem{[19]} B. Clasie et al., \emph{"Measurement of Nuclear Transparency for the A(e, e$^{'}$$\pi^{+}$) Reaction"}, Phys. Rev. Lett. 99, (2007) 242502.
\bibitem{[19A]} P. L. Jain, M. Kazuno, G. Thomas, B. Girard, Observation of nuclear transparency in p nucleus collisions at 200 GeV, Phys. Rev. Lett. 33 (1974) 660.
\bibitem{[19B]} A. I. Anoshin et al. "Investigation of $\pi^{-}$-mesons with carbon nucleus at P$\pi^{-}$=40GeV/c and the effect of "Nuclear Transparency" for high energy hadrons", Sov. Journal of Nucl. Phys. 27, (1978) 1240; Yad.Fiz. 27, (1978)1240. 
\bibitem{[20]} K. K.Gudima, V. D. Toneev., \emph{"Particle emission in light and heavy ion reactions"}, Nucl. Phys. A 400 (1983) 173.
\bibitem{[21]} A. Boudard et. al, \emph{"Intranuclear cascade model for a comprehensive description of spallation reaction data"}, Phys. Rev. C 66 (2002) 044615.
\bibitem{[23]} V. S Barashenkov et al., JINR Preprint P2-83-117, Dubna, (1983).
\bibitem{[24]} M. I. Adamovich, et. al., \emph{"Fragmentation and Multifragmentation of 10.6A GeV Gold Nuclei"}, Z. Phys. A 358 (1997) 337.
\bibitem{[25]} D. Armutlisky et al., \emph{"Multiplicity, momentum and angular distributions of protons from the interactions of p, d, á and C With carbon at 4.2-gev/c/nucleon momentum"}. Z. Phys. A 328, (1987) 455-461.
\bibitem{[26]} N. Akhababian et al. 1979 Methodical problems of determination of cross sections in inelastic interactions of relativistic nuclei with nuclei. JINR preprint 1-12114, Dubna.
\bibitem{[28]} S. Karataglidis, A. I. Wright, \emph{"Charge exchange reactions as tests for structures of exotic nuclei"} 12th International Conference on Nuclear Reaction Mechanisms, Villa Monastero, Varenna, Italy. (2009) 115.
\bibitem{[29]} Y. Afek , G. Berlad, A. Dar and G. Eilam, "Multi-pion Production in Relativistic Heavy Ion Collisions" Phys. Rev. Lett., 41, 849 (1978)
\bibitem{[30]} N. A. Kobylinsky, E.S. Martinov, V.P. Shelest, \emph{"Hadronic multiplicity and total cross-section: A new scaling in wide energy range"} Z.Phys.C-Particles and Fields 28, (1985) 143.

\end{thebibliography}
\end{document}